\input {epsf}
\documentstyle[prl,aps, multicol,epsf]{revtex}
\begin{document}
\draft
\title{The Low-Field Critical End Point of the First 
       Order Transition Line in YBa$_{2}$Cu$_{3}$O$_{7-\delta}$} 
\author{A.K.Kienappel and M.A.Moore}
\address{Department of Physics, University of Manchester,
Manchester, M13 9PL, United Kingdom.}
\date{\today}
\maketitle

\begin{abstract}
We report on simulations of the first order phase transition 
in $YBa_{2}Cu_{3}O_{7-\delta}$ using the Lawrence-Doniach model. 
We find  that the magnetization discontinuity vanishes and 
the first order transition line ends at a critical end point 
for low magnetic fields in agreement with experiment. The 
transition is not associated with  vortex lattice melting, but 
separates two vortex liquid states characterized by different 
degrees of short-range crystalline order and different length 
scales of correlations between vortices in different layers.

\end{abstract}
\begin{multicols}{2}
\narrowtext

Some features of the $B-T$  phase diagram of clean untwinned crystals of
the high temperature superconductor YBa$_{2}$Cu$_{3}$O$_{7-\delta}$
(YBCO) are widely agreed upon.
As the temperature is reduced the  vortex liquid undergoes  a first order
phase transition to what is commonly assumed to be  the Abrikosov 
vortex lattice state. 
There is striking experimental evidence for this first order transition.
Sharp drops in resistivity \cite{Ybcomelt-dyn,welp&al} as well as
discontinuities in magnetization and entropy  
\cite{Ybcomelt-stat,shilling&al} occur simultaneously and mark the
transition line. 
A  neglected feature of the experimental first order transition 
line in YBCO is its termination at 
low ($\approx 0.5T$)  magnetic fields. This 
 has been consistently observed 
in all relevant experiments (i.e. the latent heat, magnetization jump and sharp
resistance drop all disappear for  fields smaller than some lower critical
field)
\cite{Ybcomelt-dyn,welp&al,Ybcomelt-stat,shilling&al}.
Nevertheless the transition line is often shown extrapolated 
to $T_{c}$, because a transition 
which separates phases of different 
symmetries, like a vortex lattice and  a vortex liquid, 
cannot simply disappear.

\begin{figure}
\centerline{\epsfxsize= 6.7 cm\epsfbox{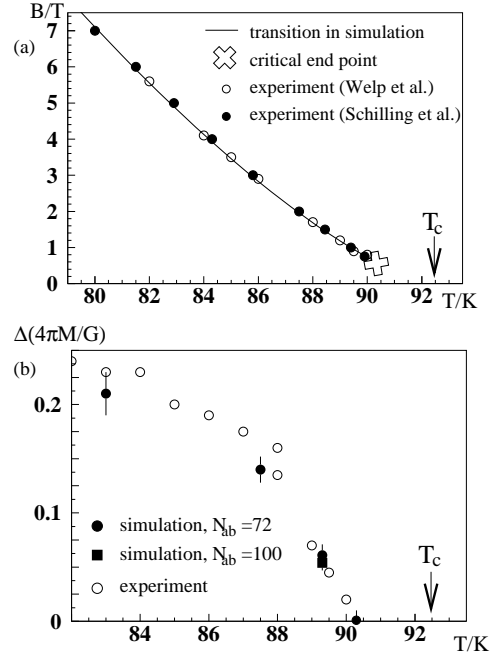}}
  \caption{Phase diagram (a)  and magnetization discontinuity 
   (b). Experimental 
   data is taken from Ref. \protect\cite{welp&al,shilling&al} (a)
   and \protect\cite{shilling&al} (b).
       $N_{ab}\equiv$ vortices per layer (for numbers of layers
       see text).}
  \label{fig:1}
\end{figure}

We report in this paper 
numerical results obtained for the Lawrence-Doniach (LD) \cite {L&D} model
 for  a clean superconductor
which show that the low field end point 
of the transition is not an artifact due to disorder (as commonly thought)
 and that the  interpretation of the first order 
transition as due to flux lattice melting must be incorrect. 
We see a first order liquid-liquid 
line with  a critical end point at low fields, which in the 
$B-T$ phase diagram  
shown in Fig.\ref{fig:1}(a) agrees very well with the experimental YBCO 
``melting'' line. On crossing  the transition line, length and time scales
 increase 
discontinuously, but remain finite even in the low-temperature phase.
The magnetization jumps we observe are 
shown in Fig.\ref{fig:1}(b) and are also in very good agreement with
measurements in YBCO. 
Figure \ref{fig:1} is obtained using standard 
YBCO values for the fitting parameters;
viz for the Landau-Ginzburg 
parameter $\kappa$=60, the mass anisotropy $\gamma$=$\sqrt{m_{c}/m_{ab}}$=7.5, 
the slope of the mean field transition line
$\partial B_{c2}/\partial T |_{T=T_{c}}=-$2T/K, 
the mean field $T_{c}$=92.5K
and the layer separation $d$=11.4\AA.  It is perhaps noteworthy 
that one of us  has frequently expressed
  doubts regarding the  
vortex-lattice melting scenario \cite{moore-PhT?&chin,moore-odlro}
on the grounds that a finite temperature 
freezing to a flux lattice state is difficult 
to reconcile with the analytical result that 
off-diagonal long range order -- a central feature
 of superconductivity --
cannot coexist with a flux lattice 
at  non-zero temperatures in three dimensions 
\cite{moore-odlro}. 

The LD model for a layered superconductor
consists of a stack of planes with Josephson coupling 
between neighboring layers.
With the superconducting order parameter in the $n^{th}$ layer denoted as 
$\psi_{n}$, the Hamiltonian for the layered system 
with a magnetic field perpendicular to the layers is
\begin{eqnarray}
&{\cal H}&=\sum_{\hbox{n}}d_{0}\int\! d^{2}\!r 
	\left(\alpha|\psi_{n}|^{2}+\frac{\beta_{2D}}{2}|\psi_{n}|^{4}+
        \right.\nonumber\\
	&& \left.\frac{1}{2m_{ab}}
               |(-i\hbar{\bf\nabla}\!-\!2e{\bf A})\psi_{n}|^2
              +J|\psi_{n+1}\!-\!\psi_{n}|^{2}
            \right),\nonumber
\end{eqnarray}
\vspace{-0.2in}
\hspace{-0.18in}
where ${\bf B}={\bf \nabla \times A}$, which we shall take as constant and
uniform.
In first approximation $\alpha(T)=\alpha '(T-T_{c})$
and $\beta_{2D}(T)$ is  constant;  $ \alpha ',\beta_{2D}, J>0$.

The model was simulated  using  Langevin (Model A) dynamics.
The  equation 
$\partial\psi/\partial t=-\Gamma \partial {\cal H}/\partial \psi^{\ast}+\eta$
was integrated numerically, using pseudo-random numbers to imitate the 
Gaussian white noise $\eta$, 
$\langle\eta^{\ast}\,\eta\rangle = 2\Gamma\, k_{B}T\,\delta_{r}\delta_{t}$. 
We studied $N_{ab}$ vortices per layer for $N_{c}$ spherical layers 
 (see Ref. \cite{dodgson&kienappel} for details of spherical boundary
conditions), each  of 
thickness $d_{0}$ and spacing $d$ and  imposed periodic boundary
 conditions
 $\psi_{1}=\psi_{N_{c}+1}$.
For each layer $\psi$ is expanded in eigenstates of the squared momentum
operator $(-i\hbar {\bf \nabla}-2e{\bf A})^{2}$. We retain eigenstates
belonging only to the
lowest eigenvalue  $2eB\hbar$, (the lowest Landau level (LLL) 
approximation)
which  is a useful procedure over a large portion of the vortex liquid 
regime \cite{Tes&Andr-ikeda}.  
The model then depends on only two dimensionless parameters 
\cite{k&m_inprep}. The first one is  
the 2D effective temperature and field parameter for each layer, given by 
$\alpha_{2T}=(d_{0}h/2e\, \beta_{2D}  B\, k_{B}T)^{1/2}\alpha_{H}$,
with $\alpha_{H}=\alpha(T)+eB\hbar/m_{ab}$.
The second, $\eta=J/|\alpha_{H}|$, controls the strength of the 
coupling between layers. 

If the order parameter varies only slowly across the layers \cite{L&D},
the layered material behaves like a continuum with a mass anisotropy $\gamma$
and $\beta=\beta_{2D}\times d/d_{0}$.
The effective mass $m_{c}$ in the $c$-direction
 can be expressed in terms of the layer 
coupling term: $\eta=\hbar^{2}\,/\,2m_{c} d^{2}|\alpha_{H}|$.
In the continuum model of an anisotropic 3D type II superconductor,
all thermodynamic properties of the sample depend on only one
parameter $\alpha_{T}\propto \alpha_{H}$. With  
$\kappa=\sqrt{\beta/2\mu_{0}}\times m_{ab}/e\hbar$
we can express $\alpha_{T}$ in terms of measurable quantities as

\vspace{-0.3in}
\begin{figure}
\centerline{\epsfxsize= 9 cm\epsfbox{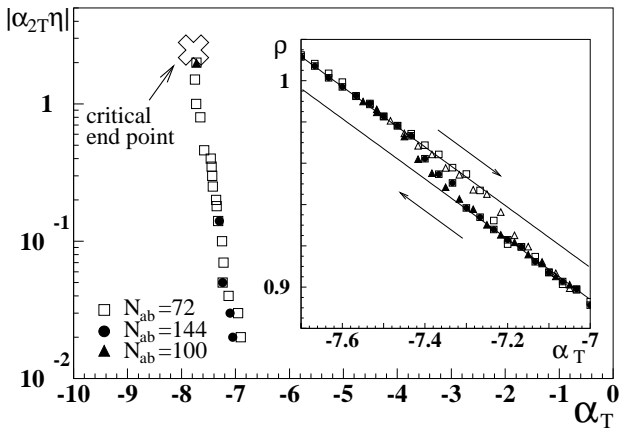}}
  \caption{First order transition points in the 
     $\alpha_{T}$--$|\alpha_{2T}\eta|$ plane. 
     $N_{c}$ varies between 8 and 80 for $|\alpha_{2T}\eta|$ 
     between 0.02 and 2. The inset 
     shows an example of the order parameter density $\rho$ at constant
    $|\alpha_{2T}\eta|$, here $|\alpha_{2T}\eta|$=0.14,
    depending on increasing and decreasing $\alpha_{T}$.
     Squares and triangles correspond to $N_{ab}$=72 
      and  $N_{ab}$=144 respectively, filled symbols to cooling
     and open symbols to heating of the system. The transition
     is marked by a clear hysteresis loop. Solid lines   
     mark the discontinuity in $\rho$. }
  \label{fig:2}
\end{figure}

\[
\alpha_{T}\!=\!\left(\frac{\sqrt{2}\hbar^{3/2}\pi}
             {k_{B} e^{3/2}\mu_{0}}\right)^{2/3}\!\! 
             \left( \frac{1}{\kappa^{2}\gamma}\right)^{2/3}
            \frac{\partial B_{c}/\partial T |_{T_{c}}(T-T_{c})+B}
                                 {(BT)^{2/3}}.
\]
Note that $\alpha_{T}$=0 corresponds to the mean-field $H_{c2}(T)$ line and
$\alpha_{T}\!=\!-\infty$ to $T$=0.
The LD Hamiltonian  reduces to the continuum Landau-Ginzburg model in the limit
$\eta \rightarrow \infty$, $\alpha_{2T}\rightarrow 0$, 
with  $\alpha_{T}^{3}= \eta (2\alpha_{2T})^{4}$ fixed.
Because YBCO is closer to being 3D rather than 2D in character,
we shall quote results in terms of $\alpha_{T}$ rather than  
$\alpha_{2T}$. 
The natural length scales of  the continuum model are the magnetic length scale
$l_{m}=\sqrt{\hbar/2eB}$, which is proportional to the vortex separation
distance in the layers, and the mean field coherence length 
$\xi_{||}=\hbar/(2m_{c}|\alpha_{H}|)^{1/2}$ parallel to the field.

Instead of $\eta$ we chose for our second independent parameter the 
product $|\alpha_{2T}\eta|$. Other 
than a factor $1/\sqrt{BT}$, $|\alpha_{2T}\eta|$ contains 
only material constants and therefore varies slowly for rather 
a wide range of $\alpha_{H}$ and over considerable regions of the $B$-$T$ 
plane.
For comparison with experiment, we can express the coupling strength as   
\[
|\alpha_{2T}\eta|=
  \left(\frac{\hbar^{3} \pi}{8 e^{3} k_{B} \mu_{0}}\right)^{1/2}
        \frac{1}{\kappa \gamma^{2} d^{3/2}}
        \frac{1}{(BT)^{1/2}}. 
\]
Note that the (unknown) 2D parameters of the  layers in the model,
$\kappa_{2D}$ 
and $d_{0}$, cancel from the the definition of $|\alpha_{2T}\eta|$
if the relation $\kappa=\sqrt{d/d_{0}} \times \kappa_{2D}$
is used. Thus we only need to know the value $\kappa$ of bulk YBCO. 

Fig. \ref{fig:2} shows the phase diagram in terms of simulation 
parameters. As  $|\alpha_{2T}\eta|$ increases
and the system approaches the continuum limit, the transition disappears 
at a critical point.  This means that a first order transition is not
expected to occur  in superconductors such as niobium for
which the continuum approximation is appropriate. 
Our simulation results imply that the end point of the first order 
transition in YBCO is intrinsically not an effect of disorder.
(Recent experiments \cite{Roulin} show that its location is significantly 
shifted to higher fields,  (which corresponds to lower coupling 
$|\alpha_{2T}\eta|$) in  twinned 
samples. This effect could be interpreted as an effective enhancement 
of the correlations along the field direction by correlated disorder. We
 expect point disorder to have only a slight effect on the location of the
end point).
Note that along the transition line $\alpha_{T}$ is approximately
constant, which means that the field and temperature dependence
 of the transition 
line behave like in a continuum model where $\alpha_{T}$ is
the only scaling parameter in the system.

The inset of Fig.\ref{fig:2} shows an example of the kind of measurement
used to locate the first order transition.
The system displays hysteresis at it  upon heating and cooling and the
 effects of
this on the order parameter density  are shown.   
The  order parameter density is given by 
$\rho=\alpha_{T}\, \beta/2 \pi \alpha_{H} \times \langle |\psi|^{2}\rangle$.
The magnetization in the LLL model is  
$4\pi M=(\mu_{0}e\hbar/m_{ab})\langle |\psi|^{2}\rangle$
(angular brackets signify a thermal average), which is in terms
of our simulation parameters 
$4\pi M=\pi(B-B_{c2}(T)) \rho/\alpha_{T}\kappa^{2}$,
where $B$ is the applied magnetic field.
Thus we can work out the magnetization discontinuity from the
discontinuity in $\rho$ at the transition.
The data points in Fig.\ref{fig:1}(b) represent  
$|\alpha_{2T}\eta|=$1, 1.5, 2 and 2.5 at $N_{c}=$50, 60, 60/80, 80.
For these 4 transition  points we have an average value of 
$\alpha_{T}= -7.72$, which yields the transition line in Fig.\ref{fig:1}(a).
For $|\alpha_{2T}\eta|$=2 and $N_{c}d/\xi_{||}=65$ we see a clear transition, 
while for $|\alpha_{2T}\eta|$=2.5 and  $N_{c}d/\xi_{||}=80$ there is no sign 
of a transition in the range $-8.3<\alpha_{T}<-7.3$. 

\begin{figure}
\centerline{\epsfxsize= 7.5 cm\epsfbox{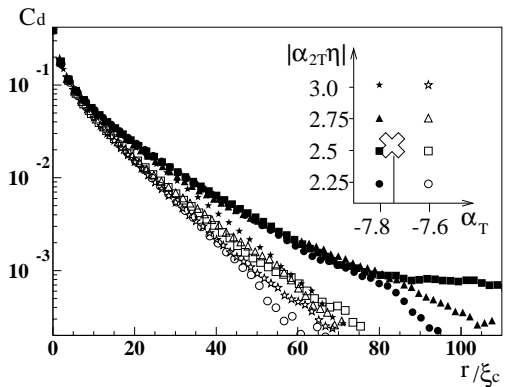}}
  \caption{Density-density correlations along the c-axis near the end 
	   the first order transition line and the critical point (see Inset). 
           Note the decreasing difference in correlations between 
           $\alpha_{T}=-7.6$ and $\alpha_{T}=-7.8$ as the transition
           disappears. For $|\alpha_{2T}\eta|=2.5$ and $\alpha_{T}=-7.8$
	   we see evidence for a long length scale associated with
	   fluctuations in the average order parameter density.
           System sizes are $N_{ab}$=72, $220<N_{c}<260$ for  
	   $\alpha_{T}=-7.6$ and  $N_{c}=270$  for $\alpha_{T}=-7.8$.
	}
  \label{fig:3}
\end{figure}

Near a critical end point we do not only expect the jump in the  
magnetization (order parameter density)
to disappear, but we also expect there to be  a divergence of
the length scale of fluctuations in the order parameter
density of the system.   
We therefore  looked at the density-density correlations of the order 
parameter near the critical point. A normalized density-density correlator 
is $C_{d}({\Delta \bf r})=\langle|\psi({\bf r})|^{2}|\psi({\bf r}+
{\Delta\bf r})|^{2}\rangle/\langle|\psi|^{2}\rangle^{2}-1.$
Let us consider the case where $\Delta \bf r$ is a vector parallel to  the
$c$-axis. 
Plots of these correlations can be seen in Fig. \ref{fig:3}. There is evidence
of two length scales in the vicinity of the end-point.
The short distance decay of the  correlation function is 
 dominated by the positional
correlations of the vortices in the different layers. This length scale
 is mostly determined by $\alpha_{T}$ and changes slowly in the vicinity of  
the critical point. However a second longer length scale
becomes visible between $\alpha_{T}=-$7.6 and $\alpha_{T}=-$7.8
 as $|\alpha_{2T}\eta|$ 
is increased to its  critical end point value. One can see in Fig. \ref{fig:3}
that when $|\alpha_{2T}\eta|=$2.5 and $\alpha_{T}=-$7.8 there is evidence of 
this much longer second length scale governing the decay of the 
correlation function at large distances. This length scale is associated 
with the density fluctuations at the critical end point and  only become 
visible once it is larger than that of the vortex 
correlations. Due to the small amplitude of these density fluctuations 

\vspace{-0.2in}

\hspace{-0.18in}
very long simulation 
times are needed to determine the correlations within the  
statistical noise.

It is often supposed that the  first order transition in YBCO
changes to second order below the end point, where
no latent heat is visible but a ``step'' in the heat capacity $C$ remains
\cite{shilling&al,Roulin}. We believe however that this ``step'' can be 
identified with the onset of a small peak in the superconducting 
specific heat $C_{s}=C-C_{n}$, (n for normal state), which 
is known to arise from thermal fluctuations. This peak  has been observed 
for example in niobium by Farrant and Gough \cite{Farrant&Gough}. 
We find that the location 
and the height of the peak as well as the  length of the rise 
(or width of the ``step'') in $\;C\;$ from  the low 
temperature value $C_{s, mf}$ 
($mf$ for mean field) to its  maximum 
agree well for the niobium and YBCO measurements taken from Ref. 
\cite{Farrant&Gough} and \cite{shilling&al}. 
The peak in  $C_{s}$ in niobium obeys LLL scaling 
\cite{Farrant&Gough} and is found at $\alpha_{T}/\beta_{A}= -6$,
where $\beta_{A}=1.16$, i.e. $\alpha_{T}\approx -7$.
The data which shows the specific heat in YBCO is given as  $C$ minus
$C(B=0)$. The latter is near the ``step'' approximately equal to the 
low temperature value $C_{s, mf}+C_{n}$, so that 
the plotted quantity is approximately $C_{s}-C_{s, mf}$. The
maximum occurs for example for $B=0.25T$  at $T\approx 91.4$K which 
corresponds to $\alpha_{T}= -7.2$.
The width of the ``step'' in niobium $\Delta \alpha_{T}\approx 2$. 
In YBCO  for  $B=0.25T$ the specific heat rise associated with the step
takes place in the temperature region  $91-91.4K$, 
which corresponds to $\Delta \alpha_{T}= 2.8$. 
For niobium  $C_{s}$ is at its maximum 5\% larger than 
$C_{s, mf}$. In YBCO we have to divide  the plotted data by 
$C_{s, mf}$ to compare with this value. $C_{s, mf}$ is roughly given 
by the step in $C$ at the zero field transition which we take from 
Ref.\cite{shilling&al}. We find that 
$(C_{s}-C_{s, mf})/C_{s, mf}\approx 0.02$, which is of the same order
as in niobium. The specific heat ``step'' in YBCO at different fields 
has approximately the same amplitude as well as width and position 
when expressed in terms of $\alpha_{T}$, 
i.e. the  ``step'' feature obeys LLL  scaling. The quantitative agreement 
between YBCO and niobium strongly suggests that 
we are dealing with the same phenomenon and therefore that there really 
is no sharp specific heat step in YBCO. 

\begin{figure}
\centerline{\epsfxsize= 9 cm\epsfbox{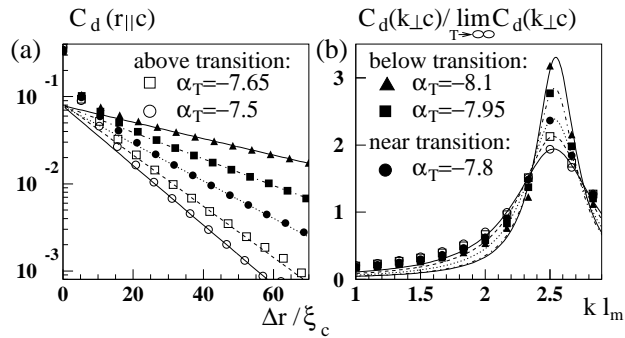}}
  \caption{(a) Density-density correlations along the c-axis 
           (with linear fits), 
           (b) structure factor (with Lorentzian fits) 
           at $|\alpha_{2T}\eta|$=1. 
           The system size is $N_{ab}$=72, $N_{c}$=270.
	}
  \label{fig:4}
\end{figure}

The existence of the critical point implies that no symmetries are  
broken at the transition, which means it cannot be a 
liquid-crystal transition.  We find indeed that the vortex matter is
liquid on both sides of the transition. In Fig. \ref{fig:4} we show
examples of 
density-density correlations above and below the transition
for $|\alpha_{2T}\eta|=1$ which corresponds to a transition temperature
of 83 K in YBCO. Fig. \ref{fig:4}(a) shows the correlations along the 
c-axis as previously seen in Fig. \ref{fig:3}.  
We see an exponential decay of the correlation function
with a finite  length scale $l_{c}$ for density correlations
below as well as above the transition.
Only for a liquid phase would this correlation function have an exponential
decay.  
In Fig. \ref{fig:4}(b) we show examples of measurements of the 2D Fourier 
transform of the density-density correlator $C_{d}$ for  ${\Delta\bf r}$ 
parallel to the layers, normalized by its high temperature 
limit, which is essentially the structure factor of the system  
(see Ref.\cite{dodgson&kienappel} for details).
It has its first and largest peak at the first 
reciprocal lattice vector of a triangular lattice, $k\approx 2.6$ in 
units of magnetic length.
The inverse width at half maximum of the Lorentzian 
fits to this peak gives the length
scale of crystalline order in the layers, $l_{ab}$.
The length scales taken from fits in figure \ref{fig:4} 
are in order of decreasing $\alpha_{T}$  
$l_{c}/\xi_{||}=$13, 15, 20, 28, 46
and  $l_{ab}/l_{m}=$1.33, 1.57, 1.94, 2.47, 2.73.

\vspace{-0.2in}

The length scales have a discontinuity at the transition.
This jump is found to grow with distance from the end point,
as one would naturally expect. 
We also see a rapid growth of the  length scales
below the transition. An exponential growth of length scales with 
$|\alpha_{T}|^{3/2}\sim (T_{c}-T)^{3/2}$,
has been predicted for the low temperature regime from 
perturbative studies around zero temperature \cite{moore-PhT?&chin}.
We can extrapolate the growth in the crystalline length 
scale obtained from Fig. \ref{fig:4} (b) just 
below the transition assuming exponential growth with 
$|\alpha_{T}|^{3/2}$.
For a decrease of $\Delta \alpha_{T} \approx 1.2$, which  corresponds 
in YBCO to cooling by only 1K below the transition at 83K, we obtain an  
increase in the range of crystalline order by a factor of 3. 
It is therefore likely that not far below the first order liquid-liquid
transition length scales reach the system size in a pure system or in real
crystals a ``Larkin" like length scale  
(dependent on the amount of disorder
present)
so that the  vortex liquid is correlated and effectively crystalline
 over large length scales. Although the structure factor in a liquid is
rotationally symmetric, coupling with the
underlying lattice may for long length scales lead to the 
appearance of Bragg--like peaks \cite{yeo}.

The longest time scales in the system, given by the decay of 
3D Fourier component of $C_{d}$ at the first reciprocal lattice vector
in the $ab$-plane and $k$=0 along the $c$-axis (not shown),
increase discontinuously across the transition. Such behavior
may  explain the sharp  features of transport coefficients 
like resistivity.  
We also see extremely fast further growth of time scales below the transition
in agreement with the fast decay of resistivity to zero as the temperature is
lowered \cite{k&m_inprep}.

From previous simulations using the same model with periodic
boundary conditions (PBC) in all directions 
instead of the geometry of spherical layers in a radial field, 
a vortex lattice melting transition is reported \cite{Sas&Str,Hu&mcD}.
We find disagreement in the location of the transition 
due to  the different choice of boundary conditions
only for low $|\alpha_{2T}\eta|$.
We believe that our choice of  boundary conditions is more likely to 
reflect the real physics \cite{bokil&moore}. 
In the 3D-like regime appropriate to YBCO, the LLL-LD model may well
show the same behavior for  PBC as it does for spherical layers. 
For couplings high enough to see the critical point, 
the LLL-LD model has to our knowledge never been investigated using PBC. 
The largest system sizes used in the simulations with PBC 
are of the order of 40 vortices $\times$ 20 layers,
smaller than the ranges of correlations we find below the transition. 
This would make the  vortex liquid indistinguishable 
from a vortex lattice. 

We have also attempted to compare our simulations with experimental data 
on Bi$_{2}$Sr$_{2}$CaCu$_{2}$O$_{8}$, for which the first order transition is 
in a region of the phase
diagram where the LLL approximation is not good. As might be
expected our  results are not as quantitative as they are for YBCO but they
are none the less still useful.
Details will be given in \cite{k&m_inprep}.

In summary, we have  found in a simulation the low field
end point of the first order transition line in YBCO and obtained results in
excellent agreement with experiment. 
The existence of a critical point implies that the vortex matter
is liquid above and below the transition, and we were able to  observe this
directly in 
our simulation. Our results 
suggest that the transition in YBCO, which is commonly interpreted 
as vortex lattice melting, is of a liquid-liquid nature and that the vortex
crystal state does not exist.

\end{multicols}

\begin{references}
\bibitem{Ybcomelt-dyn}
 A.\ Schilling  {\it et al.} , Phys.\ Rev.\ Lett \ {\bf 72}, 1092 (1994).
\bibitem{welp&al}
 U.\ Welp {\it et al.} , Phys.\ Rev.\ Lett. \ {\bf 76}, 4809 (1996);
 {\bf 67}, 3180 (1991).
\bibitem{Ybcomelt-stat}
 R.\ Liang  {\it et al.} Phys.\ Rev.\ Lett \ {\bf 76}, 853 (1996);
 M.\ Roulin, {\it et al.} Physica\ C  {\bf 275}, 245 (1997);
          Science \ {\bf 273},  1210 (1996). 
\bibitem{shilling&al}
 A.\ Schilling  {\it et al.} , Phys.\ Rev.\ Lett. \ {\bf 78}, 4833 (1997).
\bibitem{L&D}
  W.\ E.\ Lawrence and S.\ Doniach, in {\it Low Temperature Physics},
        12th international conference proceedings, Kyoto, Japan, 
        edited by E. \ Kanda, (Keygaku, Tokyo, Japan, 1971), p.316.
\bibitem{moore-PhT?&chin}
  M.\ A.\ Moore, Phys.\ Rev.\ B\ {\bf 55}, 14136 (1997);
  S-K.\ Chin and M.\ A.\ Moore, preprint cond-mat/9709347.
\bibitem{moore-odlro}
   M.\ A.\ Moore, Phys.\ Rev.\ B\ {\bf 45}, 7336 (1992).
\bibitem{dodgson&kienappel}
  M.\ J.\ W.\ Dodgson and M.\ A.\ Moore, Phys.\ Rev.\ B\ {\bf 55}, 3816 (1997),
  A.\ K.\ Kienappel and M.\ A.\ Moore, {\it ibid.} {\bf 56}, 8313 (1997).
\bibitem{Tes&Andr-ikeda}
  Z.\ Te\v{s}anovi\'{c}, A.\ V.\ Andreev, Phys.\ Rev.\ B.\ {\bf 49}, 
     4064 (1994);
  R.\ Ikeda, J.\ Phys.\ Soc.\ Jpn.\  {\bf 64}, 1683 (1995).
\bibitem{k&m_inprep}
  A.\ K.\ Kienappel and M.\ A.\ Moore, in preparation.	
\bibitem{Roulin}
   M. \ Roulin {\it et al.}, Phys. \ Rev. \  Lett. \ {\bf 80}, 1722 (1998).
\bibitem{Farrant&Gough}
 S.\ P.\ Farrant and C.\ E.\ Gough,  Phys. \ Rev. \  Lett.\ 
                   {\bf 75}, 943 (1975).
\bibitem{yeo}
 J.\ Yeo and M.\ A.\ Moore, Phys.\ Rev.\ Lett.\ {\bf 78}, 4490 (1997);
\bibitem{Sas&Str}
R.\ \v{S}\'{a}\v{s}ik and D.\ Stroud, Phys.\ Rev.\ Lett. {\bf 75}, 2582 (1995);
\bibitem{Hu&mcD}
  J.\ Hu and A.\ H.\ MacDonald, Phys.\ Rev.\ B.\ {\bf 56}, 2788 (1997).
\bibitem{bokil&moore}
  H.\ Bokil and M.\ A.\ Moore, in preparation.
\end{references}
\end{document}